\documentclass[11pt]{article}
\usepackage{latexsym,cite,amssymb}
\usepackage{times}
\makeatletter

\@addtoreset{equation}{section} \makeatother

\input amssym.def
\input amssym.tex

\newcommand{\barve}{{\bar{\varepsilon}}}

\newcommand{\G}{{\Gamma}}

\newcommand{\dis}{{\displaystyle}}
\newcommand{\pa}{{\partial}}

\newcommand{\half}{{{\textstyle\frac{1}{2}}}}

\newcommand{\be}{\begin{equation} }
\newcommand{\ee}{\end{equation} }
\newcommand{\ba}{\begin{array}}
\newcommand{\ea}{\end{array}}

\newcommand{\btheta}{\bar{\theta}}

\newcommand\New{{\scriptscriptstyle{\rm New}}}
\newcommand\aux{{\omega}}
\newcommand\NaGo{{\rm \scriptscriptstyle{N.G.}}}

\newcommand\light{{{\rm \scriptscriptstyle L.C.}}}
\def\G{{g}}

\def\cG{{\cal G}}
\def\cL{{\cal L}}
\def\cM{{\cal M}}

\def\cS{{\cal S}}

\def\rd{{\rm d}}
\newcommand\hd{\hat{d}}
\def\WZ{{{\scriptscriptstyle{\rm WZ}}}}
\def\NB{{{\scriptscriptstyle{\rm N.B}}}}
\def\PB{{{\scriptscriptstyle{\rm P.B}}}}

\def\Tr{{\rm Tr}}
\def\I_N{{1_{\scriptscriptstyle N\times N}}}

\begin{document}
\begin{titlepage}
\title{
\vskip 2cm
Three-algebra for supermembrane and two-algebra for superstring\\~\\}
\author{\sc
Kanghoon Lee${}^{\sharp}$ \mbox{~~\,\,}and\mbox{\,\,~~} Jeong-Hyuck Park${}^{\dagger}$}
\date{}
\maketitle \vspace{-1.0cm}
\begin{center}
~~~\\
${}^{\sharp}$Department of Physics,   Yonsei University, Shinchon-dong, Seodaemun-gu, Seoul 120-749, Korea\\
\texttt{lkh@phya.yonsei.ac.kr}\\
~{}\\
${}^{\dagger}$Department of Physics, Sogang University, Shinsu-dong, Mapo-gu, Seoul 121-742, Korea\\
\texttt{park@sogang.ac.kr}
~{}\\
~~~\\
~~~\\
\end{center}
\begin{abstract}
\vskip0.5cm
\noindent
While string or Yang-Mills theories are based on Lie algebra or two-algebra structure,  recent studies  indicate  that $\cM$-theory may require a one higher,  three-algebra structure. Here we construct  a covariant  action for a supermembrane in eleven dimensions,  which is invariant under   global supersymmetry, local fermionic symmetry and  worldvolume diffeomorphism. Our action is classically on-shell equivalent to the celebrated  Bergshoeff-Sezgin-Townsend action. However, the novelty is that
we spell the action genuinely in terms of Nambu three-brackets: All the derivatives appear through Nambu brackets and hence  it manifests  the three-algebra structure. Further the double dimensional reduction of our action  gives straightforwardly to  a  type IIA string action featuring  two-algebra. Applying the same method, we also construct a covariant action for type IIB superstring,  leading directly  to the  IKKT matrix model.

\end{abstract}
\thispagestyle{empty}
\end{titlepage}
\newpage

\tableofcontents 
\section{Introduction: Two for string and three for $\cM$-theory}
While   string and Yang-Mills theories are based on ordinary Lie-algebra or two-algebra structure,   recent advances in $\cM$-theory  by
Bagger, Lambert and Gustavsson (BLG)~\cite{BL,Gustavsson:2007vu}   suggest  that the full description of $\cM$-theory may require a generalized  Lie-algebra structure: namely   three-Lie algebra or shortly three-algebra.  In fact,      the digits, two and  three, appear to  have  intriguing associations   to
string and $\cM$-theory respectively: First of all,  two is the dimension of string worldsheet while three is that of membrane worldvolume. This implies that,  after  matrix regularization of Poisson bracket structure,  IKKT matrix model~\cite{Ishibashi:1996xs} is a multiple D-instanton description of  type IIB superstring  \textit{via} two-algebra, while
BFSS matrix model~\cite{Banks:1996vh,deWit:1988ig} is a multiple  D$0$-brane description of  eleven-dimensional  supermembrane \textit{via} two-algebra.  In other words, 
the two worldsheet coordinates of a type IIB superstring are traded with matrix indices, while the three-dimensional  worldvolume of a supermembrane  decomposes into `1+2', one for the temporal coordinate and two for the matrix indices. Further, two is the codimension  of D-branes in each type IIA, IIB superstring theory~\cite{Polchinski:1998rr}, while three is the codimension of M-branes \textit{i.e.~}M2 and M5. Consequently, through  two-algebra interaction as known as Myers effect~\cite{Myers:1999ps},  multiple D$p$-branes may condense or be polarized into D${(p+2)}$-brane. Similarly, through three-algebra interaction, BLG model equipped with an infinite dimensional gauge group  corresponds to a description of  the condensation  of multiple M2-branes into a single M5-brane~\cite{Gomis:2008cv,Ho:2008nn,Ho:2008ve,
Krishnan:2008zm,bps1,Park:2008qe,Bandos:2008fr,Bonelli:2008kh,Bandos:2008jv,bps2,Lee:2008cr,Ganjali:2008bq,Ganjali:2009kt}.  Namely, the polarizations of D-branes and M2-branes require two-algebra and three-algebra respectively.

\begin{itemize}
\item{2 for string}:\\
\begin{tabular}{cll}
String worldsheet dimension &$\Longrightarrow$& IKKT type IIB matrix model \\
2-algebra structure &$\Longrightarrow$& Matrix string\,/\,Yang-Mills\\
Codimension of D-branes~~&$\Longrightarrow$~~&  Myers effect: polarization  of D$p$  into D${(p+2)}$\\

\end{tabular}

\item{3 for $\cM$-theory}:\\
\begin{tabular}{cll}
Membrane worldvolume dimension  &$\Longrightarrow$& BFSS $\cM$-theory matrix model  \\
3-algebra structure &$\Longrightarrow$& BLG model\\
Codimension of M2 and M5&~~$\Longrightarrow$~~&Condensation of M2s  into M5 in BLG model\\

\end{tabular}

\end{itemize}
~\\

All the above associations of 2 and 3 to string and $\cM$-theory may be naturally understood by a  reformulation of the Nambu-Goto action for a $p$-brane. Prior to the explanation,   we first review Filippov  $n$-Lie algebra and discuss its  generalization which is necessary for us later.

\subsection{Filippov $n$-Lie algebra and its generalization  \label{Lie}}
Filippov introduced  $n$-Lie algebra~\cite{n-Lie} which is a natural generalization of a Lie algebra,  defined by   $n$-bracket satisfying the totally anti-symmetric property:
\be
[X_{1},\cdots,X_{i},\cdots,X_{j},\cdots,X_{n}]=-[X_{1},\cdots,X_{j},\cdots,X_{i},\cdots,X_{n}]\,,
\label{antisym}
\ee
and the Leibniz rule:
\be
\left[X_{1},\cdots,X_{{n-1}},[Y_{1},\cdots,Y_{n}]\right]=\sum_{j=1}^{n}~
\left[Y_{1},\cdots,[X_{1},\cdots,X_{{n-1}},Y_{j}],\cdots,Y_{n}\right]\,.
\label{Leibniz}
\ee
The $n$-Lie algebra can be equipped with an invariant inner product, satisfying the symmetric property,
\be
\langle X, Y\rangle=\langle Y, X\rangle\,,
\ee
as well as  the invariance under the  $n$-bracket transformation,
\be
\langle [X_{1},\cdots,X_{{n-1}},Y], Z\rangle+\langle Y,[X_{1},\cdots,X_{{n-1}},Z]\rangle=0\,.
\label{invbra}
\ee
When $n =2$ the definition reduces to  the  usual Lie algebra and the inner product can be given  by  `Trace'.

Explicitly we may introduce a basis of the $n$-Lie algebra, $T^{a}$, $a=1,2,\cdots$ and write
\be
[T^{a_{1}},T^{a_{2}},\cdots,T^{a_{n}}]=f^{a_{1}a_{2}\cdots a_{n}}{}_{b}T^{b}\,.
\ee
From (\ref{antisym}) the structure constant $f^{a_{1}a_{2}\cdots a_{n}}{}_{b}$ is totally anti-symmetric for the upper indices, and the Leibniz rule implies
\be
f^{a_{1}a_{2}\cdots a_{n}}{}_{c}f^{b_{1}b_{2}\cdots b_{n}}{}_{a_{n}}=\sum_{j=1}^{n}~
f^{a_{1}a_{2}\cdots a_{{n{-}1}}b_{j}}{}_{d}\,f^{b_{1}\cdots b_{{j{-}1}}db_{{j{+}1}}\cdots b_{n}}{}_{c}\,.
\ee
Further the invariant inner product defines a metric $\langle T^{a},T^{b}\rangle$ which, along with its inverse, can raise or lower the index $a$. It is worth while to note that in the above expressions all the quantities are assumed to be bosonic. When   fermionic variables are present, there must appear extra minus sign if the fermionic quantities are permuted odd times.

One may easily  realize the $n$-Lie algebra in terms of  Nambu $n$-bracket defined over  functional space on an   $n$-dimensional manifold~\cite{Nambu:1973qe}:
\be
\ba{cll}
[X_{1},X_{2},\cdots,X_{n}]~&\Longleftrightarrow&~
\{X_{1},X_{2},\cdots,X_{n}\}_{\NB}:=
\textstyle{\frac{1}{\sqrt{{\cG}}}}\epsilon^{l_{1}l_{2}\cdots l_{n}}
\partial_{l_{1}}X_{1}\partial_{l_{2}}X_{2}\cdots\partial_{l_{n}}X_{n}\,,\\
\langle X,Y\rangle~&\Longleftrightarrow&~
\displaystyle{\int}{\rm d}^{n}y\,\sqrt{{\cG\,}}  XY\,.
\ea
\label{NBG}
\ee

In order to ensure the partial integration, either the manifold must be compact or
all the functions must vanish on the boundaries of the non-compact manifold. Note  that $\cG$ corresponds to the determinant of the metric of the manifold, and   can be chosen  arbitrarily  since the properties (\ref{antisym}), (\ref{Leibniz}), (\ref{invbra})  hold irrespective of the presence of the local factor.
In this functional realization of the $n$-Lie algebra,  the invariant inner product generalizes:
\be
\ba{lll}
\langle X,Y,\cdots,Z\rangle~&\Longleftrightarrow&~
\displaystyle{\int}{\rm d}^{n}y\sqrt{{\cG}} XY\cdots Z\,,
\ea
\label{generalXYZ}
\ee
such that it satisfies, as a generalization of (\ref{invbra}),
\be
\sum_{k=1}^{m}~
\langle Y_{1},Y_{2},\cdots,Y_{k-1},[X_{1},
\cdots,X_{{n-1}},Y_{k}],Y_{{k+1}},\cdots\,Y_{m}\rangle=0\,.
\label{invbra2}
\ee

Note that throughout  the paper we denote the defining equality by `$:=$' and  the on-shell equality as well as   gauge fixings   by  `$\equiv$'.  For further works on three-algebra see \textit{e.g.} \cite{Papadopoulos:2008sk,Gauntlett:2008uf,Axenides:2006ms,Axenides:2008rn}.

\subsection{Reformulation of   Nambu-Goto action  by Nambu bracket}
With an embedding of  ${({p+1})}$-dimensional worldvolume coordinates  into  $D$-dimensional target spacetime,
\be
X(\xi)~:~\xi^{m}~\longrightarrow~X^{M}\,,
\ee
where $m=0,1,\cdots, p~$ and $M=0,1,\cdots,{D-1}$, the Nambu-Goto action for a $p$-brane reads~\cite{NambuGoto}
\be
S_{\NaGo}=\displaystyle{-\int}\rd^{p{+1}}\xi~\sqrt{-
\det(\partial_{m}X^{M}\partial_{n}X_{M})\,}\,.
\label{NGp}
\ee
Let us decompose, formally, the $p$-brane  worldvolume coordinates  into two parts:
\be
\{\,\xi^{m}\,\}=\{\,\sigma^{\mu}\,,\,\varsigma^{i}\}\,,
\ee
where  $\mu=0,1,\cdots, {d-1}~$ and $i=1,\cdots,\hd~$ such that $p+1=d+\hd$.
The decomposition  is \textit{ a priori}
arbitrary  for any non-negative integers $d,\hd$. One natural  application of the splitting will be the case where $p$-brane is extended over two topologically different spaces, \textit{e.g.~}compact and non-compact spaces. With the decomposition above, a square root free reformulation of the Nambu-Goto action was achieved in \cite{Park:2008qe}:
\be
\ba{l}
{S=\displaystyle{{\int}}\rd^{d}\sigma~\Tr\left(\sqrt{-h\,}L\,\right)\,,}~~~~~~~~~~~~
\dis{\Tr:=\int\rd^{\hd}\varsigma\,,}\\
L=-h^{\mu\nu}D_{\mu}X^{M}D_{\nu}X_{M}
-\textstyle{\frac{1}{\,4{\hd!\,}}}\aux^{d-1}
\{X^{M_{1}},X^{M_{2}},\cdot\cdot,X^{M_{\hd}}\}_{\NB}
\{X_{M_{1}},X_{M_{2}},\cdot\cdot,X_{M_{\hd}}\}_{\NB}+(d-1)\aux\,,
\ea
\label{actionNew}
\ee
where the action contains three kinds of  auxiliary fields: scalar $\aux$, $d$-dimensional metric  $h_{\mu\nu}$ and a gauge connection $A_{\mu}^{~i}$ which defines the  `covariant derivative':
\be
D_{\mu}X^{M}:=\partial_{\mu}X^{M}-A_{\mu}^{~i}\partial_{i}X^{M}\,.
\ee
The Nambu $\hd$-bracket~(\ref{NBG})  is defined here, simply  without a local factor,  by\footnote{As usual, $\epsilon^{i_{1}i_{2}\cdots i_{{\hd}}}$ is   the totally anti-symmetric $\hd$-dimensional  tensor of the normalization  $\epsilon^{12\cdots{\hd}}=1$.}
\be
\{Y_{1},Y_{2},\cdots,Y_{\hd}\}_{\NB}
:=\epsilon^{i_{1}i_{2}\cdots i_{{\hd}}}\partial_{i_{1}}Y_{1}\partial_{i_{2}}Y_{2}\cdot\cdot
\partial_{i_{\hd}}Y_{\hd}\,.
\label{NambuB}
\ee
Integrating out all the auxiliary fields, using their on-shell values, the action reduces to the Nambu-Goto action, $S_{\New}\equiv S_{\NaGo}$, and hence the classical equivalence.  The novelty of the above reformulation was the appearance  of the \textit{gauge interaction}   and the  \textit{Nambu bracket} squared potential.
The latter basically stems  from an identity  rewriting  the determinant as  the  Nambu bracket squared:
\be
\det(\partial_{i}X^{M}\partial_{j}X_{M})=
\textstyle{\frac{1}{\hd!}}\{X^{M_{1}},X^{M_{2}},\cdot\cdot,X^{M_{\hd}}\}_{\NB}
\{X_{M_{1}},X_{M_{2}},\cdot\cdot,X_{M_{\hd}}\}_{\NB}\,.
\label{identity2}
\ee

 A physical picture behind  the reformulation is to describe a single brane as a condensation of multiple lower-dimensional branes,  \textit{i.e.~} a  $p$-brane by $(d-1)$-branes:  IKKT matrix model~\cite{Ishibashi:1996xs} is a multiple D-instanton description of  type IIB superstring, while  BFSS matrix model~\cite{Banks:1996vh} is a multiple  D$0$-brane description of   supermembrane (see also \cite{Yang:1998qd}) . Obviously, the choice of $\hd=0$ and $d={p+1}$ corresponds  to the well-known    ``Polyakov" action which was actually first conceived by
Brink, Di Vecchia, Howe, Tucker~\cite{Brink:1976sc,Howe:1977hp}. On the other hand,   with a gauge fixing for $\omega$ to be constant, the other extreme choice of $d=0$, $\hd={p+1}$  leads to the Schild action~\cite{Schild:1976vq}. Furthermore, the association of the digits, $2$ and $3$ to string and $\cM$-theory  become manifest    within the reformulation: For example, the fact that the codimension of D-branes is 2 suggests to choose $\hd=2$, which leads to the two-algebra as in Yang-Mills. Likely the choice of $p=5$, $d=3$, $\hd=3$ suggests that the Bagger-Lambert-Gustavsson model with an infinite dimensional gauge group describes a M5-brane as a condensation of multiple M2-branes.

The reformulation of the Nambu-Goto action (\ref{actionNew})   is purely bosonic. In order to establish a firm connection to string/$\cM$-theory one needs to supersymmetrize them.  The requirement of  supersymmetry may give rise to a constraint on the  \textit{a priori} arbitrary decomposition,  $p+1=d+\hd$.

Our main interest is to  supersymmetrize  the action (\ref{actionNew}).
For $d=1$ case,  supersymmetric actions are   ready to be read-off from an earlier work
by Bergshoeff, Sezgin, Tanii and Townsend~\cite{Bergshoeff:1988hw}.
In its appendix the authors listed  light-cone gauge fixed  supersymmetric actions for  various $p$-branes in diverse spacetime dimensions. Utilizing the identity (\ref{identity2}), in terms of Nambu $p$-bracket, their light-cone gauge fixed supersymmetric $p$-brane  actions can be reexpressed  in a compact form:
\be
\cL_{\light}=\half (D_{t}X^{I})^{2}-\textstyle{\frac{1}{2p!}}
\{X^{I_{1}},X^{I_{2}},\cdot\cdot,X^{I_{p}}\}_{\NB}^{2}+i\half\bar{\Psi}D_{t}\Psi+
\textstyle{\frac{1}{2(p-1)!}}\bar{\Psi}\Gamma^{I_{1}I_{2}\cdots I_{p-1}}\{X_{I_{1}},\cdots,X_{I_{p-1}},\Psi\}_{\NB}\,.
\ee
As usual, the Fierz identity required  for the supersymmetry invariance, restricts the possible values of $p$ and the spacetime dimension $D$ (as for $I=1,2,\cdots,D-2$):
\be
\ba{ll}
p=1~:&~D=3,4,6,10\\
p=2~:&~D=4,5,7,11\\
p=3~:&~D=6,8\\
p=4~:&~D=9\\
p=5~:&~D=10\,.
\ea
\ee

In the present paper, we consider an alternative choice of $d=0$. In particular,  we focus  on a supermembrane propagating in eleven-dimensional flat spacetime. As we take the choice of $\hd=3$,  the bosonic action is of Schild type and it will contain manifestly Nambu three-brackets.\\
~\\

The organization of the rest of the paper is as follows. In section~\ref{M2} we present  our main result: We  construct  a  3-algebra based  action for a supermembrane in eleven dimensions. The action is invariant under   global supersymmetry, local kappa-symmetry and  worldvolume diffeomorphism. In section~\ref{IIA} we perform a double dimensional reduction and obtain a covariant, two-algebra based  action for a type IIA superstring  in ten dimensions. In a similar fashion, in section~\ref{IIB} we construct a covariant, two-algebra based  action for a type IIB superstring. Section~\ref{DISCUSS} contains  our discussion and the appendix carries some useful identities.\\

\textit{Note added}: While this work was being finished, two related papers,  \cite{Sato:2009mf} and \cite{Furuuchi:2009ax}, appeared on arXiv. The former discusses a bosonic  three-algebra squared action and the latter presents a supersymmetric version of it with the  spacetime dimension four and the signature $2+2$. On the other hand, our action is for the supermembrane in eleven-dimensional Minkowskian spacetime.

\newpage
\section{Three-algebra based action for  supermembrane in eleven dimensions \label{M2}}
\subsection{The action}
We propose the following action for the three-algebra description of a supermembrane in eleven dimensions:
\be
\ba{ll}
\cS_{\rm M2} = \displaystyle\int \rd^3\xi \left(\cL_{\omega} + \cL_{\WZ}\right)\,,
\ea
\label{auxm2action0}
\ee
\be
\label{3lag1}
\ba{l}
\cL_{\omega}= {1\over12}\omega^{-1} \langle\Pi^{M},\Pi^{N},\Pi^{P}\rangle\langle\Pi_{M},\Pi_{N},\Pi_{P} \rangle -{1\over2}\omega\,,\\
\cL_{\WZ} = -i{1\over2}\epsilon^{ijk}
\btheta\Gamma_{MN}\pa_{i}\theta\left(\Pi^{M}_{j}\pa_k X^{N}-{1\over3}\btheta\Gamma^{M}\pa_j\theta \,\btheta\Gamma^{N}\pa_k\theta\right)\,,
\ea
\ee
which contains  eleven-dimensional target spacetime  coordinates $X^{M}$,  ~a  Majorana spinor $\theta$ and a  scalar density field $\omega$. The former two are dynamical while the last one is auxiliary.  With the  supersymmetry invariant pull-back:
\be
\Pi^{M}_{i} := \pa_{i} X^{M} - i \btheta \Gamma^{M}\pa_{i}\theta\,,
\ee
we set
\be
\langle \Pi^{L},\Pi^{M},\Pi^{N}\rangle:=\epsilon^{ijk}\Pi_{i}^{L}\Pi_{j}^{M}\Pi_{k}^{N}\,,
\ee
which has the following expansion in terms of  the Nambu-bracket (\ref{NambuB}),\footnote{Note that the bracket $[L,M,N]$ denotes the anti-symmetrization  of the three indices with an overall factor $\frac{1}{6}$, \textit{i.e} of `strength one'.}
\be
\ba{lll}
\langle \Pi^{L},\Pi^{M},\Pi^{N}\rangle&:=&\{X^{L},X^{M},X^{N}\}_{\NB}-3i\bar{\theta}\Gamma^{[L}\{X^{M},X^{N]},\theta\}_{\NB}
+3\btheta\{\Gamma^{[L}\theta,X^{M},\btheta \Gamma^{N]}\}_{\NB}\theta
\\
&{}&-i\btheta_{\alpha}\btheta_{\beta}\btheta_{\gamma}
\{(\Gamma^{[L}\theta)^{\alpha},(\Gamma^{M}\theta)^{\beta},(\Gamma^{N]}\theta)^{\gamma}\}_{\NB}\,.
\ea
\label{PPPNB}
\ee
Similarly, the Wess-Zumino part of the action can be also reexpressed  in terms of the  Nambu-bracket:
\be
\ba{lll}
\cL_{\WZ} &=& -i\half \btheta \Gamma_{MN}\{X^{M},X^{N},\theta\}_{\NB}+\half \btheta_{\alpha}\btheta_{\beta}\{(\Gamma_{MN}\theta)^{\alpha},(\Gamma^{M}\theta)^{\beta},X^{N}\}_{\NB}
\\
&{}& -i{1\over 6}\btheta_{\alpha}\btheta_{\beta}\btheta_{\gamma}
\{(\Gamma_{MN}\theta)^{\alpha},(\Gamma^{M}\theta)^{\beta},
(\Gamma^{N}\theta)^{\gamma}\}_{\NB}\,.
\ea
\ee
Thus, all the derivatives appear only through Nambu three-brackets.\\

Let us  now introduce a shorthand notation for  the induced metric:
\be
\G_{ij} := \Pi_{i}^{M} \Pi_{Mj}\,,
\ee
and denote its determinant by $\G:=\det(\G_{ij})$ as usual.

All the equations of motion are then summarized by:
\be
\ba{l}
\omega-\sqrt{-\G}\equiv0\,,\\
\G^{ij}\Pi^{M}_{i}\Gamma_{M}(1-\Gamma)\partial_{j}\theta\equiv0\,,\\
\partial_{i}\left(\sqrt{-\G}\G^{ij}\Pi_{j}^{M}\right)-i\epsilon^{ijk}\partial_{i}\btheta
\Gamma^{M}{}_{N}\partial_{j}\theta\Pi^{N}_{k}\equiv0\,.
\ea
\label{M2eom}
\ee

From an identity  analogue  to (\ref{identity2}):
\be
\textstyle{\frac{1}{6}} \langle\Pi^{M},\Pi^{N},\Pi^{P}\rangle\langle\Pi_{M},\Pi_{N},\Pi_{P} \rangle=\det\left(\Pi_{i}^{M}\Pi_{jM}\right)\,,
\ee
integrating out the auxiliary scalar  assuming  the on-shell value  $\aux\equiv\sqrt{-g}$, our proposed action (\ref{auxm2action0}) reduces to the well-known supersymmetric Nambu-Goto action for M2-brane by Bergshoeff, Sezgin and Townsend~\cite{Bergshoeff:1987cm,Bergshoeff:1987qx}:
\be
\label{m2action}
\displaystyle{
{\cS}_{\rm M2} \equiv \int \rd^3 \xi \Big[-\sqrt{-\det\left(\Pi_{i}^{M}\Pi_{jM}\right)} \,-i\textstyle{1\over2}\epsilon^{ijk}
\btheta\Gamma_{MN}\pa_{i}\theta\left(\Pi^{M}_{j}\pa_k X^{N}-\textstyle{1\over3}\btheta\Gamma^{M}\pa_j\theta \,\btheta\Gamma^{N}\pa_k\theta\right)\Big]\,.}
\ee


\subsection{Symmetries}
The action (\ref{auxm2action0}) is invariant under the following transformations.
\begin{itemize}
\item \textit{Target-spacetime supersymmetry}:
\be
\ba{lll}
\delta_{\varepsilon} \theta = \varepsilon\,,~~~~&~~~~
\delta_{\varepsilon} X^{M} = -i \btheta \Gamma^{M}\varepsilon\,, ~~~~&~~~~
\delta_{\varepsilon}\omega=0\,,
\ea
\ee
which leaves $\Pi^{M}_{i}$ and $\cL_{\omega}$ invariant, while transforms $\cL_{\WZ}$ to a total derivative:
\be
\delta_{\varepsilon} \cL_{\WZ} =\pa_{i}(\barve \eta^{i})\,,
\ee
where $\varepsilon$ is a constant supersymmetry parameter and
\be
\ba{ll}
\eta^{i} :=& -i\half \epsilon^{ijk} \Big\{\,\Gamma_{MN}\theta \,\pa_{j}X^{M}\pa_{k}X^{N}
-i\left(\Gamma_{MN} \pa_{j} \theta\, \btheta \Gamma^{M} \pa_{k} \theta -\Gamma^{M} \pa_{k}\theta\,\btheta\Gamma_{MN} \pa_{j} \theta\,  \right)X^{N}\\
&~~~~~~~~~~~~~~~~~~-{1\over15}\left(\Gamma_{MN}\theta\,\btheta \Gamma^{M}\pa_{j}\theta + \Gamma^{M}\theta \,\btheta \Gamma_{MN} \pa_{j} \theta\right)\btheta\Gamma^{N}\pa_{k}\theta \Big\}\,.
\ea
\ee
~\\

\item \textit{Local 32-component fermionic symmetry}:
\be
\ba{l}
\delta_{\zeta}\theta=\left[1+(\textstyle{{\omega/\sqrt{-\G}}})\Gamma\right]\zeta\,,\\
\delta_{\zeta} X^{M} = i\btheta \Gamma^{M}\delta_{\zeta}\theta\,,\\
\delta_{\zeta} \omega =
4i\omega\G^{-1ij}\Pi^{M}_{i}\pa_{j}\btheta\Gamma_{M}\zeta\,,
\ea
\label{zeta}
\ee
where $\zeta$ is an arbitrary local 32-component   spinorial   parameter and $\Gamma$ is  as in \cite{Bergshoeff:1987cm,Bergshoeff:1987qx}:
\be
\textstyle{\Gamma:= {1\over6\sqrt{-\G}}\,\Gamma_{LMN}\langle\Pi^{L},\Pi^{M},\Pi^{N}\rangle\,,}
\label{w/oauxGamma}
\ee
satisfying
\be
\Gamma^{2} = 1\,.
\ee
Under the  transformation above (\ref{zeta}), the  Lagrangian  transforms to a total derivative:
\be
\delta_{\zeta}\left(\cL_{\omega}+\cL_{\WZ}\right) =  \pa_{i}\left(\bar{\psi^{i}}\delta_{\zeta}\theta\right)\,,
\ee
where
\be
\bar{\psi^{i}}:= -\half \epsilon^{ijk}\Big[i\pa_{j}X^{M}\pa_{k}X^{N}\btheta\Gamma_{MN}
+ \left(\btheta\Gamma^{M}\pa_{j}\theta\, \btheta\Gamma_{MN}
+\btheta\Gamma_{MN}\pa_{j}\theta\,\btheta\Gamma^{M} \right)\!
\left(\pa_{k}X^{N}-i\textstyle{1\over3}\btheta\Gamma^N \pa_{k}\theta\right)\Big]\,.
\ee

In particular, taking the choice $~\zeta=(1+\omega/{\sqrt{-\G}})^{-1}(1+\Gamma)\kappa~$ leads to a
symmetry:
\be
\ba{l}
\delta_{\kappa}\theta = (1 +\Gamma)\kappa\,,\\
\delta_{\kappa} X^{M} = i\btheta \Gamma^{M}\delta_{\kappa}\theta\,,\\
\delta_{\kappa} \omega = 4i{\omega\sqrt{-\G}\over \omega+\sqrt{-\G}}\,\G^{-1ij}\Pi^{N}_{i}\pa_{j}\btheta\Gamma_{N}\delta_{\kappa}\theta\,,
\ea
\label{kappa}
\ee
where $\kappa$ is an arbitrary local fermionic parameter so that the transformations of $\theta$ and $X^{M}$ coincide with the kappa-symmetry in \cite{Bergshoeff:1987cm,Bergshoeff:1987qx}.

On the other  hand, an  alternative, in fact complimentary, choice $~\zeta=(1+\omega/{\sqrt{-\G}})^{-1}[1-(\omega/{\sqrt{-\G}})\Gamma\,]\kappa^{\prime}~$ leads to a   symmetry:
\be
\ba{l}
\delta_{\kappa^{\prime}}\theta = (1 -\omega/{\sqrt{-\G}})\kappa^{\prime}\,,\\
\delta_{\kappa^{\prime}} X^{M} = i\btheta \Gamma^{M}\delta_{\kappa^{\prime}}\theta\,,\\
\delta_{\kappa^{\prime}} \omega = 4i{\omega\sqrt{-\G}\over \omega+\sqrt{-\G}}\,\G^{-1ij}\Pi^{N}_{i}\pa_{j}\btheta\Gamma_{N}
[1-(\omega/{\sqrt{-\G}})\Gamma\,]\kappa^{\prime}\,.
\ea
\label{kappaprime}
\ee
On-shell (\ref{M2eom}), these transformations  are trivial and hence cannot  be used to reduce the fermionic physical degrees further after a $\kappa$-gauge fixing. More discussion on trivial symmetry  transformations we refer \textit{e.g.} \cite{Henneaux:1992ig}. Combining (\ref{kappa}) and (\ref{kappaprime}) gives back the generic transformation (\ref{zeta}), and hence the former two are complimentary to each other.

\item \textit{Worldvolume diffeomorphism}:
\be
\ba{lll}
\delta_{v} X^{M} = v^{i}\partial_{i}X^{M}\,,~~~~&~~~~
\delta_{v} \theta =  v^{i}\partial_{i}\theta\,,~~~~&~~~~
\delta_{v} \omega =  \partial_{i}(\omega v^{i})\,,
\ea
\ee
where $v^{i}=\delta\xi^{i}$ is an arbitrary local bosonic parameter,  and the Lagrangian transforms to a total derivative as
\be
\ba{ll}
\delta_{v}\cL_{\omega}=\partial_{i}(v^{i}\cL_{\omega})\,,~~~~~&~~~~
\delta_{v}\cL_{\WZ}=\partial_{i}(v^{i}\cL_{\WZ})\,.
\ea
\label{diffvL}
\ee
\end{itemize}

\newpage

\section{Double dimensional reduction to Type IIA superstring theory\label{IIA}}
Double dimensional reduction~\cite{Duff:1987bx} of our supermembrane action~(\ref{auxm2action0}), putting $\xi^{2}\equiv X^{10}$, $\Gamma^{(11)}:=\Gamma^{10}$,   straightforwardly leads to the following reformulation of the  type IIA superstring action by Green and Schwarz~\cite{Green:1983wt, Green:1983sg}:\footnote{For the dimensional reduction of BLG model see \textit{e.g.~}\cite{Santos:2008ue,Franche:2008hr}. }
\be
\cS_{\rm IIA} = \int \rd^2 \xi (\cL_{\omega} + \cL_{\WZ})\,,
\label{IIAS}
\ee
where with $i=1,2$, $M=0,1,\cdots,9$,
\be
\ba{cll}
\cL_{\omega} &=& {1\over4} \omega^{-1} \langle\Pi^M , \Pi^N \rangle \langle\Pi_M , \Pi_N \rangle -\half \omega\,,
\\
\cL_{\WZ} &=& i\epsilon^{ij}\pa_{i}X^{M} \btheta\Gamma_{M(11)}\pa_{j}\theta -\half \epsilon^{ij} \btheta \Gamma_{M(11)} \pa_{i}\theta \,\btheta\Gamma^{M} \pa_{j}\theta\,,
\ea
\label{IIAaction}
\ee
and
\be
\ba{ll}
\Pi^{M}_{i} = \pa_{i}X^{M} -i \btheta \Gamma^{M} \pa_{i} \theta\,,~~~~&~~~~
\langle\Pi^{M} , \Pi^{N} \rangle := \epsilon^{ij}\Pi^{M}_{i}\Pi^{N}_{j}\,.
\ea
\ee
Note that $\theta$ is a ten-dimensional Majorana spinor which can decompose into a chiral  and an anti-chiral Majorana spinor, and hence type IIA Majorana-Weyl spinors of opposite chiralities.

In terms of  Nambu 2-bracket or Poisson bracket, we can write
\be
\ba{l}
\langle\Pi^{M} , \Pi^{N} \rangle= \{X^{M},X^{N}\}_{\PB} -2i\btheta\{X^{[M},\Gamma^{N]}\theta\}_{\PB}
+\btheta_{\alpha}\btheta_{\beta}
\{(\Gamma^{[M}\theta)^{\alpha},(\Gamma^{N]}\theta)^{\beta}\}_{\PB}\,,\\
\cL_{\WZ} = i \btheta\{X^{M},\Gamma_{M (11)}\theta\}_{\PB} +\half \btheta_{\alpha}\btheta_{\beta} \{(\Gamma_{M (11)}\theta)^{\alpha}, (\Gamma^{M} \theta)^{\beta}\}_{\PB}\,.
\ea
\ee
Thus, all the derivatives appear only through Poisson brackets.

Integrating out the auxiliary scalar field $\omega$, our action~(\ref{IIAS}) reduces to the Green-Schwarz action for type IIA
superstring~\cite{Green:1983wt,Green:1983sg}.\\

The  action~(\ref{IIAS})  is invariant under the following transformations.
\begin{itemize}
\item \textit{Target-spacetime supersymmetry}:
\be
\ba{lll}
\delta_{\varepsilon} \theta = \varepsilon\,,~~~~&~~~~
\delta_{\varepsilon} X^{M} = -i \btheta \Gamma^{M}\varepsilon\,, ~~~~&~~~~
\delta_{\varepsilon}\omega=0\,,
\ea
\ee
which leaves $\cL_{\omega}$ invariant and transforms $\cL_{\WZ}$ to a total derivative:
\be
\delta_{\varepsilon} \cL_{\WZ} =\pa_{i}(\barve \eta^{i}_{\scriptscriptstyle{\rm IIA}})\,,
\ee
where $\varepsilon$ is a constant supersymmetry parameter and
\be
\eta^{i}_{\scriptscriptstyle{\rm IIA}} := \epsilon^{ij} \Big[-i\Gamma_{M(11)}\theta\, \pa_{j}X^{M} -\textstyle{1\over6}\Gamma_{M(11)}\theta \,\btheta\Gamma^{M} \pa_{j}\theta-\textstyle{1\over6}\Gamma^{M}\theta \,\btheta\Gamma_{M(11)} \pa_{j}\theta\,\Big]\,.
\ee

\item \textit{Local 32-component fermionic symmetry}:
\be
\ba{l}
\delta_{\zeta} \theta =[1+(\omega/\sqrt{-g})\tilde{\Gamma}\Gamma^{(11)}]\zeta\,,
\\
\delta_{\zeta} X^{M} = i\btheta \Gamma^{M} \delta_{\zeta}\theta\,,
\\
\delta_{\zeta} \omega = 4i\omega g^{-1 ij}\Pi^{M}_{i}\pa_{j}\btheta \Gamma_{M} \zeta\,,
\ea
\label{zetaIIA}
\ee
where $\zeta$ is an arbitrary 32-component local fermionic parameter and
\be
\tilde{\Gamma}:= \textstyle{{1\over2\sqrt{-\G}}}\, \epsilon^{ij} \Pi^{M}_{i}\Pi^{N}_{j}\Gamma_{MN}\,,
\ee
satisfying $\tilde{\Gamma}^2=1$. Under the transformation~(\ref{zetaIIA})  the Lagrangian transforms to a total derivative:
\be
\delta_{\zeta}\left(\cL_{\omega}+\cL_{\WZ}\right) =  \pa_{i}\left(\bar{\psi}_{\scriptscriptstyle{\rm IIA}}^{i}\delta_{\zeta}\theta\right)\,,
\ee
where
\be
\bar{\psi}_{\scriptscriptstyle{\rm IIA}}^{i} :=\epsilon^{ij}\left(-i\pa_{j}X^{M} \btheta\Gamma_{M(11)} -\half \btheta\Gamma_{M(11)}\pa_{j}\theta \,\btheta\Gamma^{M} -\half \btheta\Gamma^{M}\pa_{j}\,\btheta\Gamma_{M(11)}\right)\,.
\ee
As in the case of   the three-algebra based supermembrane action (\ref{kappa}), (\ref{kappaprime}), the  local fermionic symmetry consists of  $\kappa$-symmetry and trivial transformation.

\item \textit{Worldvolume diffeomorphism}:
\be
\ba{lll}
\delta_{v} X^{M} = v^{i}\partial_{i}X^{M}\,,~~~~&~~~~
\delta_{v} \theta =  v^{i}\partial_{i}\theta\,,~~~~&~~~~
\delta_{v} \omega =  \partial_{i}(\omega v^{i})\,.
\ea
\ee
The Lagrangian transforms to a total derivative as (\ref{diffvL}).\\
\end{itemize}

\section{Type IIB superstring theory and IKKT matrix model\label{IIB}}
In a similar fashion to our type IIA superstring action~(\ref{IIAS}), the
Schild version of type IIB superstring action assumes the form:
\be
\cS_{\rm IIB} = \int \rd^2 \xi (\cL_{\omega} + \cL_{\WZ}) \,,
\label{IIBS}
\ee
where
\be
\ba{cll}
\cL_{\omega} &=& {1\over4} \omega^{-1} \langle\Pi^M , \Pi^N \rangle \langle\Pi_M , \Pi_N \rangle -\half \omega\,,
\\
\cL_{\WZ} &=& -i\epsilon^{ij}\pa_{i}X^{M}(\btheta^{1}\Gamma_{M}\pa_{j}\theta^{1}-\btheta^{2}\Gamma_{M}\pa_{j}\theta^{2}) +\epsilon^{ij}\btheta^{1}\Gamma^{M}\pa_{i}\theta^{1}\,\btheta^{2}\Gamma_{M}\pa_{j}\theta^{2}\\
&=& -i\btheta^{1}\Gamma_{M}\{X^{M},\theta^{1}\}_{\PB}
+i\btheta^{2}\Gamma_{M}\{X^{M},\theta^{2}\}_{\PB} -\btheta^{1}_{\alpha}\btheta^{2}_{\beta}
\{(\Gamma_{M}\theta^{1})^{\alpha},(\Gamma^{M}\theta^{2})^{\beta}\}_{\PB}\,.
\ea
\ee
With a pair of Majorana-Weyl spinors of a same chirality,
\be
\ba{ll}
\Gamma^{(11)} \theta^{1} = \theta^{1}\,, ~~~~&~~~~
\Gamma^{(11)} \theta^{2} = \theta^{2}\,,
\ea
\ee
the supersymmetry invariant pull-back is given by
\be
\Pi^{M}_{i} = \pa_{i}X^{M} -i\left(\btheta^{1} \Gamma^{M} \pa_{i} \theta^{1} +\btheta^{2} \Gamma^{M} \pa_{i} \theta^{2} \right)\,.
\ee
~\\

Our type IIB superstring  action (\ref{IIBS}) is invariant under the following transformations.
\begin{itemize}
\item \textit{Target-spacetime supersymmetry}:
\be
\ba{llll}
\delta_{\varepsilon} \theta^{1} = \varepsilon^{1}\,,~~~~&~~~~
\delta_{\varepsilon} \theta^{2} = \varepsilon^{2}\,,~~~~&~~~~
\delta_{\varepsilon} X^{M} = -i\btheta^{1}\Gamma^{M}\varepsilon^{1}
-i\btheta^{2}\Gamma^{M}\varepsilon^{2}\,, ~~~~&~~~~
\delta_{\varepsilon}\omega=0\,,
\ea
\label{IIBve}
\ee
which leaves $\cL_{\omega}$ invariant and transforms $\cL_{\WZ}$ to a total derivative:
\be
\delta_{\varepsilon} \cL_{\WZ} =\pa_{i}( \barve^{1}\eta^{1 i}_{\scriptscriptstyle{\rm IIB}}+\barve^{2}\eta^{2 i}_{\scriptscriptstyle{\rm IIB}})\,,
\ee
where $\varepsilon^{1}$, $\varepsilon^{2}$ are constant supersymmetry parameters  and
\be
\ba{ll}
\eta^{1i}_{\scriptscriptstyle{\rm IIB}}:= ~\epsilon^{ij} \left(i \pa_{j}X^{M}\Gamma_{M}\theta^{1} +\textstyle{1\over3}\Gamma^{M}\theta^{1} \,\btheta^{1}\Gamma_{M} \pa_{j}\theta^{1} \right)\,,\\
\eta^{2i}_{\scriptscriptstyle{\rm IIB}}:= -\epsilon^{ij} \left(i \pa_{j}X^{M}\Gamma_{M}\theta^{2} +\textstyle{1\over3}\Gamma^{M}\theta^{2}
\,\btheta^{2}\Gamma_{M} \pa_{j}\theta^{2} \right)\,.
\ea
\ee

\item \textit{Local 32-component fermionic symmetry}:\\
\be
\ba{l}
\delta_{\zeta} \theta^{1} = (1+{\omega\over\sqrt{-g}}\tilde{\Gamma})\zeta^{1}\,,
\\
\delta_{\zeta} \theta^{2} = (1-{\omega\over\sqrt{-g}}\tilde{\Gamma})\zeta^{2}\,,
\\
\delta_{\zeta} X^{M} = i\btheta^{1} \Gamma^{M} \delta_{\zeta}\theta^{1}
+i\btheta^{2} \Gamma^{M} \delta_{\zeta}\theta^{2}\,,
\\
\delta_{\zeta} \omega = 4i\omega g^{-1ij}\Pi^{N}_{i}(\pa_{j}\btheta^{1}\Gamma_{N}\zeta^{1}
+\pa_{j}\btheta^{2}\Gamma_{N}\zeta^{2})\,,
\ea
\label{kappaIIB}
\ee
where $\zeta$ is an arbitrary local fermionic parameter. The  Lagrangian  transforms to a total derivative,
\be
\delta_{\zeta}\left(\cL_{\omega}+\cL_{\WZ}\right) =  \pa_{i}\left(\bar{\psi}_{\scriptscriptstyle{\rm IIB}}^{1 i}\delta_{\zeta}\theta^{1} + \bar{\psi}_{\scriptscriptstyle{\rm IIB}}^{2 i}\delta_{\zeta}\theta^{2}\right)\,,
\ee
where
\be
\ba{l}
\bar{\psi}_{\scriptscriptstyle{\rm IIB}}^{1 i}:=\epsilon^{ij}\left( i \Pi^{\mu}_{j} \btheta^{1} \Gamma_{M}
-\btheta^{2}\Gamma^{M}\pa_{j}\theta^{2}\,\btheta^{1}\Gamma_{M}
\right)\,,\\
\bar{\psi}_{\scriptscriptstyle{\rm IIB}}^{2 i}:=-\epsilon^{ij}\left(i\Pi^{\mu}_{j} \btheta^{2}\Gamma_{M}
-\btheta^{1}\Gamma^{M}\pa_{j}\theta^{1}\,\btheta^{2}\Gamma_{M} \right)\,.
\ea
\ee
The local fermionic symmetry above  consists of  $\kappa$-symmetry and trivial transformation, as in the supermembrane and type IIA superstring cases.

\item \textit{Worldvolume diffeomorphism}:
\be
\ba{llll}
\delta_{v} X^{M} = v^{i}\partial_{i}X^{M}\,,~~~~&~~~~
\delta_{v} \theta^{1} =  v^{i}\partial_{i}\theta^{1}\,,~~~~&~~~~
\delta_{v} \theta^{2} =  v^{i}\partial_{i}\theta^{2}\,,~~~~&~~~~
\delta_{v} \omega =  \partial_{i}(\omega v^{i})\,,
\ea
\ee
where $v^{i}=\delta\xi^{i}$ is an arbitrary local bosonic parameter. The Lagrangian transforms to a total derivative as (\ref{diffvL}).\\
\end{itemize}

Now, replacing $\theta^{2}$ by $i\theta^{2}$  through an analytic continuation~\cite{Ishibashi:1996xs}, fixing  the gauge $\theta^{1}\equiv\theta^{2}$ through the local fermionic symmetry and $\omega\equiv 1$ through the diffeomorphism, the action reduces  to the form:
\be
\cS_{\rm IIB} \equiv \displaystyle{\int \rd^2 \xi~}
 \textstyle{\frac{1}{4}}\{X^{M},X^{N}\}_{\PB}\{X_{M},X_{N}\}_{\PB}
  -2i\btheta^{1}\Gamma_{M}\{X^{M},\theta^{1}\}_{\PB}\,,
\label{IIBSIKKT}
\ee
which straightforwardly   leads, after a matrix regularization of the Poisson bracket, to the type IIB  IKKT matrix model~\cite{Ishibashi:1996xs}. Thus, our two-algebra based covariant type IIB superstring action~(\ref{IIBS}) presents a direct derivation of the IKKT matrix model starting from the covariant superstring action (\ref{IIBS}).\footnote{Note that in the original derivation of the type IIB matrix model~\cite{Ishibashi:1996xs}, the auxiliary scalar field vanished away  during the matrix regularization as it is absorbed into either the trace or  the matrix commutator, while in our scheme it is gauge fixed to be a constant.}\\

\section{Discussion\label{DISCUSS}}
In summary, we have constructed  covariant  actions  for  type IIA, type IIB superstrings in ten dimensions, and supermembrane in eleven dimensions, which
are invariant under   global supersymmetry, local fermionic symmetry and  worldvolume diffeomorphism.  All the derivatives therein appear through Nambu  brackets such that     the two-algebra structure of  superstring theory and  the three-algebra structure of   $\cM$-theory become manifest.  Nambu two and three brackets naturally arise since the dimensions of the string worldsheet and the membrane worldvolume are two and three respectively.
 One advantage to employ the Nambu brackets is the simplicity of the double dimensional reduction: The three-bracket clearly reduces to the two-bracket.

Since our resulting actions (\ref{auxm2action0}), (\ref{IIAS}), (\ref{IIBS}) contain higher than second   order terms, the generalization of the inner product of Filippov $n$-Lie algebra as in (\ref{generalXYZ}), (\ref{invbra2}) is necessary.  Like the type IIB case, suitable gauge fixing to simplify the actions for supermembrane and type IIA superstring is desired.

The BFSS matrix model~\cite{Banks:1996vh,deWit:1988ig} is a light-cone gauge fixed action for supermembrane such that it describes only the sector of classically fixed light-cone momentum.   Our covariant supermembrane action  (\ref{auxm2action0}) is classically equivalent to the BFSS matrix model. However, the quantum equivalence is to be investigated in future work. 

~\\
\textbf{Acknowledgments}\\
We wish to thank Kazuyuki Furuuchi, Seungjoon Hyun,  Corneliu Sochichiu for useful comments, and especially  Johanna Erdmenger  at Max-Planck Institut fur Physik, Munchen for helpful discussion as well as hospitality during our visit.
The work is in part supported  by the Korea Foundation for International Cooperation of Science \& Technology with grant number K20821000003-08B1200-00310,  by   the Center for Quantum Spacetime of Sogang University with grant number R11-2005-021,   by the Korea Science and Engineering Foundation with grant number R01-2007-000-20062-0,  and by the Basic Research Program of the Korea Science and Engineering Foundation with grant number R01-2004-000-10651-0.

\newpage

\appendix
\section{Useful  Fierz identities in eleven dimensions}
In Minkowskian eleven dimensions,
 with an anti-symmetric charge conjugation matrix $C$,
the gamma matrices satisfy
\be
\left(C\Gamma^{M_{1}M_{2}\cdots M_{n}}\right)^{T}=(-1)^{1+\half n(n+1)}C\Gamma^{M_{1}M_{2}\cdots M_{n}}\,,
\ee
and a  Fierz identity:
\be
(C\Gamma^{M})_{(\alpha\beta}(C\Gamma_{MN})_{\gamma\delta)} = 0\,.
\ee
Thanks to this Fierz identity, various identities follow
which are  crucial for the supersymmetry and the  local fermionic symmetry  of the action (\ref{auxm2action0}):
\be
\ba{ll}
&\epsilon^{ijk}\Big(\barve\Gamma_{MN}\theta\pa_{j}\,\btheta\Gamma^{M}\pa_{k}\theta -2\btheta\Gamma_{MN}\pa_{j}\theta  \,\pa_{k}\btheta\Gamma^{M}\varepsilon +\pa_{j}\btheta\Gamma_{MN}\pa_{k}\theta\, \barve\Gamma^{M}\theta -2\barve\Gamma_{MN}\pa_{j}\theta\,\btheta\Gamma^{M}\pa_{k}\theta  \Big)=0\,,
\\
&\epsilon^{ijk}\Big(\barve\Gamma_{MN}\pa_{i}\theta\,\pa_{j}\btheta\Gamma^{M}\pa_{k}\theta + \pa_{i}\btheta\Gamma_{MN}\pa_{j}\theta \, \barve\Gamma^{M}\pa_{k}\theta\Big)=0\,,
\\
&\epsilon^{ijk}\Big(\barve\Gamma_{MN}\pa_{i}\theta \,\btheta\Gamma^{M}\pa_{j}\theta + \btheta\Gamma_{MN}\pa_{i}\theta\,\pa_{j}\btheta\Gamma^{M}\varepsilon \Big)\btheta\Gamma^{N}\pa_{k}\theta\\
& ~~~~~~~~~~~~~~~~~~~~~~~~~~~~~~~~~~~~~~~~~~~~~~~~~~~~~= {1\over5}\epsilon^{ijk}\pa_{i}\left\{\Big(\barve\Gamma_{MN}\theta \, \btheta\Gamma^{M}\pa_{j}\theta + \btheta\Gamma_{MN}\pa_{j}\theta\barve\Gamma^{M}\theta\Big) \btheta\Gamma^{N}\pa_{k}\theta\right\}\,.
\ea
\ee

Other useful identities include
\be
\ba{cll}
(\Gamma^{M})^{[\alpha}_{~\gamma}(\Gamma_{M})^{\beta]}_{~\delta}&=& {11\over32}C^{-1\alpha\beta}C_{\gamma\delta}
+{5\over 32\times 3!}(\Gamma^{LMN}C^{-1})^{\alpha\beta}(\Gamma_{LMN}C)_{\gamma\delta}
\\
&&+{3\over32\times 4!}(\Gamma^{LMNP}C^{-1})^{\alpha\beta}(\Gamma_{LMNP}C)_{\gamma\delta}\,,
\\
(\Gamma^{MN})^{[\alpha}_{~(\gamma}(\Gamma_{N})^{\beta]}_{~\delta)}
&=& {5\over16}C^{-1\alpha\beta}(C\Gamma^{M})_{\gamma\delta} + {3\over32}(\Gamma^{ML_{1}L_{2}}C^{-1})^{\alpha\beta}(C\Gamma_{L_{1}L_{2}})_{\gamma\delta}
\\
&&+{1\over16\times 4!}(\Gamma_{L_{1}L_{2}L_{3}L_{4}}C^{-1})^{\alpha\beta}(C\Gamma^{ML_{1}L_{2}L_{3}L_{4}})_{\gamma\delta}\,,
\ea
\ee
and
\be
\sqrt{-\G} \G^{-1 ij} \Pi^{M}_{j}\Gamma_{M}\Gamma = \sqrt{-\G} \G^{-1 ij} \Pi^{M}_{j} \Gamma\Gamma_{M} = \half \epsilon^{ijk} \Pi^{L}_{j}\Pi^{M}_{k}\Gamma_{LM}\,.
\ee


\end{document}